\documentclass[%
 onecolumn, superscriptaddress,
showpacs,preprintnumbers,
nofootinbib,
 amsmath,amssymb,
 aps,
pra,
]{revtex4}
\usepackage{mathrsfs}
\usepackage{graphicx}
\usepackage{epstopdf}
\usepackage{dcolumn}
\usepackage{bm}
\usepackage{color} 
\usepackage{CJK}

\usepackage{physics}

\def\la{{\langle}}
\def\ra{{\rangle}}

\newcommand{\beq}{\begin{equation}}
\newcommand{\eeq}{\end{equation}}
\newcommand{\beqa}{\begin{eqnarray}}
\newcommand{\eeqa}{\end{eqnarray}}

\begin{document}

\title{Shortcut-to-adiabaticity-like techniques for parameter estimation in quantum metrology}

\author{Marina Cabedo-Olaya}
\affiliation{Departamento de Qu\'\i mica F\'\i sica, UPV/EHU, Apdo 644 Bilbao 48080, Spain}
\author{J. Gonzalo Muga}
\affiliation{Departamento de Qu\'\i mica F\'\i sica, UPV/EHU, Apdo 644 Bilbao 48080, Spain}
\author{Sof\'\i a Mart\'\i nez-Garaot}
\affiliation{Departamento de Qu\'\i mica F\'\i sica, UPV/EHU, Apdo 644 Bilbao 48080, Spain}

\begin{abstract}
Quantum metrology makes use of  quantum mechanics to improve precision measurements and measurement sensitivities. It is usually formulated for time-independent Hamiltonians, but time-dependent Hamiltonians may offer advantages, such as a $T^4$ time dependence of the Fisher information which cannot be reached  with a time-independent Hamiltonian. 
In \textit{Optimal adaptive control for quantum metrology with time-dependent Hamiltonians}  {(Nature Communications 8, 2017), Shengshi Pang and Andrew N. Jordan} put forward a  Shortcut-to-adiabaticity (STA)-like  method, {{specifically}} an approach 
formally similar to the ``counterdiabatic approach'', 
adding a control term to the original Hamiltonian  to reach the upper bound of the Fisher information. We revisit this work 
from the point of view of STA to set the relations and differences between STA-like methods in metrology 
and ordinary STA. 
This analysis paves the way for the application of other STA-like techniques in parameter estimation. 
In particular we explore the use of physical unitary transformations to propose alternative time-dependent 
Hamiltonians which may be easier to implement  in the laboratory. 
\end{abstract}
\maketitle

\section{Introduction}
\label{Chapter:Introduction}
Quantum metrology aims at high-resolution and highly sensitive measurements of parameters using advantages provided by quantum states and dynamics. 
Most of the research in this field has been focused on time-independent Hamiltonians, but time-dependent Hamiltonians may beat  precision limits found for time-independent ones \cite{ArticlePang,Naghiloo2017,Yang2017,Gefen2017,Mukherjee2019}
to estimate some parameter $g$ in the Hamiltonian. 

The Cram\`{e}r--Rao bound states  that the mean squared deviation  $\langle \delta^2 g \rangle$ for an unbiased 
estimation is bounded as
\begin{equation}
\label{CRB}
\langle \delta^2 g \rangle \geq \frac{1}{N I_g},
\end{equation} 
where $N$ is a measure of the amount of data and $I_g$ is the Fisher Information. 

In a quantum scenario the information about a  parameter $g$ in the Hamiltonian $H_g$ is ``stored'' in the quantum states of the system
$|\psi_g(t)\ra$, whose evolution depends on $g$, and the Fisher information for a final time $T$ measures the distinguishability (distance) between $|\psi_g(T)\ra$ and $|\psi_{g+\delta g}(T)\ra$.  The ``maximum''
Fisher information  {{for}} $g$  with respect to all possible quantum measurements is \cite{BraunsteinStatisticaldistance,BraunsteinUncertaintyrelations} 
\begin{equation}
\label{qi}
I_g^{(Q)} = 4 \left( \bra{\partial_g \psi_g (T)}\ket{ \partial_g \psi_g(T)}-|\bra{ \psi_g(T)}\ket{ \partial_g \psi_g(T)}|^2\right),
\end{equation}
which is also named ``quantum Fisher information''. 

For a given Hamiltonian $H_g$, the quantum Fisher information has an upper bound,   
\begin{equation}
I_g^{(Q)}\leq \left[\int_0^T (\mu_{max}(t)-\mu_{min}(t))dt\right]^2,
\label{ub}
\end{equation}
where $\mu_{max}(t)$ and $\mu_{min}(t)$ are the (instantaneous) maximum and minimum eigenvalues of $\partial_g H_g(t)$. 
To actually implement  this upper bound with particular states and measurements, the dynamics should follow some specific path, along an equal superposition of the corresponding eigenvectors.   
Reaching the upper bound of the Fisher information may require Hamiltonian control \cite{ArticlePang}, i.e., adding an extra term $H_c(t)$ to the original Hamiltonian of the system $H_g(t)$ to implement the necessary dynamics.   
This methodology based on driving the system along preselected  ``rails'' (states), is formally quite similar to the one proposed in Shortcut-to-adiabaticity (STA)  methods \cite{ReviewSTA}, specifically in the  ``counterdiabatic'' (CD) approach   \cite{Unanyan1997,Demirplak2003,Demirplak2005,Demirplak2008,Berry2009,Chen2010_123003}.
In the CD approach, an auxiliary Hamiltonian $H_{cd}(t)$ is also added to some reference Hamiltonian $H_0(t)$ to drive the system along eigenstates of $H_0(t)$.   
We shall revisit the main concepts and results in \cite{ArticlePang} in Sections  \ref{sec:pang} and \ref{chap:estimation},  and analyze in detail the relations and differences between ``actual STA'' and the STA-like method used in metrology, see Table \ref{table1} for an overview.  
A recurring 
topic within the counterdiabatic approach is that it is often difficult to implement in practice $H_{cd}$ \cite{Muga2010,ReviewSTA,Torrontegui2013}. This problem has lead to a number of approximations, variational approaches, or  methods based on unitary transformations that, properly adjusted, 
could be as well applicable in metrology. In Section \ref{Chapters:Unitary}, we explore in particular  the use of alternative Hamiltonians to $H_g+H_c$ 
via unitary transformations.    In the final discussion, we shall comment on prospects to apply other STA-like approaches.

\begin{table}[t]
\centering
\begin{tabular}{c|c|c|}
\cline{2-3}
& \textbf{STA}                                                                                       & \textbf{Metrology}                                                                                               \\ \hline
\multicolumn{1}{|c|}{$\ket{\psi_k(t)}$}                                                             & \begin{tabular}[c]{@{}c@{}}Eigenstates of a \\ reference Hamiltonian $H_0(t)$\end{tabular}            & \begin{tabular}[c]{@{}c@{}}Eigenstates of\\  $\partial_gH_g(t)$\end{tabular}                                     \\ \hline
\multicolumn{1}{|c|}{\begin{tabular}[c]{@{}c@{}}Reference Hamiltonian\\  $H_{ref}=H_0, H_g$\end{tabular}} & \begin{tabular}[c]{@{}c@{}}$H_0$ diagonal in\\  the basis $\{\ket{\psi_k}\}$\end{tabular}              & \begin{tabular}[c]{@{}c@{}}$H_g$ not diagonal in\\  the basis $\{\ket{\psi_k}\}$\end{tabular}                        \\ \hline
\multicolumn{1}{|c|}{$H_{cd}$}                                                                   & Drives along adiabatic states of $H_0$                                                                     & \begin{tabular}[c]{@{}c@{}}Drives along eigenstates  of $\partial_gH_g(t)$\\ (so ``CD" is here an abuse of language)\end{tabular}                        \\ \hline
\multicolumn{1}{|c|}{$H_{ref}+H_{cd}$}                                                           & \begin{tabular}[c]{@{}c@{}}The addition of $H_0$ \\ only changes phases $\theta_k(T)$\end{tabular} & \begin{tabular}[c]{@{}c@{}}Adding $H_g$ would produce transitions\\ so it is added AND subtracted, see (\ref{htot})\end{tabular} \\ \hline
\multicolumn{1}{|c|}{Speed}                                                                      & Emphasis on Fast driving                                                                           & Not necessarily fast                                                                                             \\ \hline
\multicolumn{1}{|c|}{Iterations}                                                                      
&\begin{tabular}[c]{@{}c@{}} ``Superadiabatic",\\ structural changes\end{tabular}
&\begin{tabular}[c]{@{}c@{}} Adaptive,\\ only the parameter changes\end{tabular}                                                                                              \\ \hline
\end{tabular}
\caption{Main differences between the use of STA-like methodology in metrology, following ref. \cite{ArticlePang},
and ordinary applications of STA (Counterdiabatic approach).
\label{table1}}
\end{table}


%
%
%
%
\section{Optimal Adaptive Control for Quantum Metrology with Time-Dependent Hamiltonians}
\label{sec:pang}
Our first objective is to summarize and comment on the work in \cite{ArticlePang} to set and understand the relations and differences 
between the STA-like approach applied there and ordinary STA.  The analysis should be useful for  a practitioner of STA methods less acquainted with quantum metrology, as well as for quantum metrologists not aware of the rich toolbox of  STA techniques. In the following, $\hbar=1$.   

\subsection*{Quantum Fisher Information}
\label{subsec:qparameterestimation}

The quantum Fisher information in Equation (\ref{qi}) can be rewritten as (proportional to) a variance 
computed for the initial state $\ket{\psi_0}$,
\begin{equation}
 \label{FI}
  I_g^{(Q)}=4 {\rm Var}[h_g(T)]_{\ket{\psi_0}}, 
 \end{equation}
 where  
 \begin{equation}
 \label{hg}
 h_g(T)=i ~ U^{\dagger}_g(0\rightarrow T)~ \partial_g U_g(0\rightarrow T), 
 \end{equation}
and $U_g(0\to T)$ is the unitary evolution operator from $0$ to time $T$ for the  Hamiltonian $H_g(t)$.
Being a variance, an ``optimal'' value of the quantum Fisher information,  with respect to all possible initial states, is 
\begin{equation}
\label{IQtau}
	I_{g~~op}^{(Q)}(T)=\left[\tau_{max}(T)-\tau_{min}(T)\right]^2,
\end{equation}
where $\tau_{max}(T)$ and $\tau_{min}(T)$ are maximal and minimal eigenvalues of $h_g(T)$, respectively. 
The optimal state $|\psi_0\ra$ is an equal superposition of the maximal and minimal eigenvalues of $h_g(T)$, and  the calculation of $I_{g~~op}^{(Q)}(T)$ requires diagonalizing $h_g(T)$. 
A mathematical upper bound  for this optimal value may be found by 
rewritting $h_g(T)$ in integral form \cite{ArticlePang}, 
\begin{equation}
 \label{hgint}
 h_g(T)=\int_{0}^{T}U_g^{\dagger}(0 \rightarrow t)\partial_g H_g(t) U_g(0 \rightarrow t) dt.
\end{equation}
The variance would be maximized by maximizing the contribution of each instant $t$ by means of a 
hypothetical dynamical state that were at all times an equal superposition of the eigenvectors of $\partial_g H_g(t)$ with maximal and minimal instantaneous eigenvalues $\mu_{max}(t)$, $\mu_{min}(t)$,    
\begin{equation}
\label{IQub}
I^{(Q)}_{g~~ub}=\left[\int_{0}^{T} (\mu_{max}(t)-\mu_{min}(t))dt\right]^2.
\end{equation}
We have termed this upper bound as ``mathematical'' because the eigenvectors $|\psi_{min}(0)\ra$ or $|\psi_{max}(0)\ra$ of $\partial_gH_g(0)$, 
with $\mu_{min}(0)$ and $\mu_{max}(0)$ eigenvalues, will not be driven in general along corresponding eigenvectors  $|\psi_{min}(t)\ra$ and  $|\psi_{max}(t)\ra$ with eigenvalues $\mu_{min}(t)$ and $\mu_{max}(t)$ by $H_g(t)$,  
i.e., in general the optimal value does not reach (saturate) the upper bound,  
\begin{eqnarray}
\label{maxtau}
\tau_{max}(T) \leq \int_{0}^{T}\mu_{max}(t)dt,\;\;\;
\tau_{min}(T) \geq \int_{0}^{T}\mu_{min}(t)dt.
\end{eqnarray}
It is important to distinguish the  quantum Fisher information in (\ref{FI}) (which is ``maximal'' with respect to measurements, for a given state), the ``optimal'' quantum Fisher information (\ref{IQtau}) (with respect to measurements and states), and the ``upper bound'' (\ref{IQub}). 
The optimal value can be calculated in principle from the Hamiltonian $H_g(t)$ alone, but to implement  it in an actual estimation protocol  
we would need specific states and measurements. Similarly, the upper bound depends formally only on $H_g(t)$ 
(more specifically on its derivative $\partial_gH_g(t)$), but its realization also needs careful state and measurement 
selection, as well as extra control terms, as we shall see.    
The terms ``maximal'', ``optimal'', and ``upper bound'' referred to the Fisher information represent here {{an}} ordered hierarchy but could be confusing and  are subjected to specific defining conditions, 
they have to be put in context. It is not easy to keep an entirely consistent terminology, for example, ``an optimal value''
or an ``upper bound'' are of course also maximal values in some sense. Moreover  Pang and Jordan refer to the process that achieves the upper bound (\ref{IQub}) by adding Hamiltonian control as   ``optimal'' \cite{ArticlePang}.
Even the concept of an ``upper bound'' is a relative one as it depends on the chosen $H_g$. In summary, dealing with  
this somewhat entangled  parlance needs careful reading. 

Equation (\ref{hgint}) gives the clue to physically realize the upper bound. Again, the initial state must be an equal combination of the 
maximal and minimal eigenvectors of $\partial_g H_g(0)$,  and their time evolution should keep them as instantaneous 
eigenvectors of $\partial_g H_g(t)$ until $t=T$.   
The solution proposed in \cite{ArticlePang} to achieve this guided  dynamics  is to add a control term $H_c$ to the Hamiltonian $H_g$ so that the states are driven by  the total Hamiltonian in the same way instantaneous eigenstates $\ket{\psi_k(t)}$ of $\partial_g H_g(t)$ change with time,  
\begin{equation}
\label{htotal}
	H_{tot}(t)=H_g(t)+H_c(t).
\end{equation}
Here is where the core similarity with STA (counterdiabatic) methods lays. Both in counterdiabatic methods and 
in the parameter estimation strategy set in \cite{ArticlePang}, new terms are added to some 
reference Hamiltonian so that the system is guided along predetermined paths. 

The proposed form of the control term is  \cite{ArticlePang}
\begin{equation}
\label{hc}
H_c(t)=\sum_{k} f_k(t) \ket{\psi_k(t)} \bra{\psi_k(t)}-H_g(t)+i \sum_{k}\ket{\partial_t \psi_k(t)}\bra{\psi_k(t)}, 
\end{equation}
where the $f_k(t)$ are in principle arbitrary functions of time that could be chosen for convenience, and 
the $|\psi_k(t)\ra$ are the instantaneous eigenvectors  of $\partial_g H_g(t)$. 
Rewriting $H_c$ as 
\begin{equation}
\label{HcHcd}
H_c(t)=-H_{g}(t)+H_{cd}(t),
\end{equation}
where
\begin{equation}
\label{generalHcd}
H_{cd}(t)=\sum_{k} f_k(t) \ket{\psi_k(t)} \bra{\psi_k(t)}+i \sum_{k}\ket{\partial_t \psi_k(t)}\bra{\psi_k(t)},
\end{equation}
then 
\beq
\label{first}
H_{tot}(t)=H_{cd}(t).
\eeq
Equation (\ref{generalHcd}) may be found by {\it imposing} a unitary evolution operator of the form 
$U(0\to t)=\sum_k e^{-i\theta_k(t)} |\psi_k(t)\ra\la \psi_k(0)|$, which drives the dynamics along the eigenstates of $\partial_g H_g(t)$
up to phase factors $e^{-i\theta_k(t)}$.   
The corresponding Hamiltonian must be $i\dot{U}U^\dagger$  
which gives exactly the right hand side in Equation (\ref{generalHcd}) with      
\begin{equation}
\label{theta}
\theta_k(t)=\int_{0}^{t} f_k(t')dt'.
\end{equation}
In STA applications, $H_{cd}(t)$ 
is called counterdiabatic Hamiltonian \cite{ReviewSTA} because, in that context, it avoids diabatic transitions among eigenstates of $H_0(t)$. 
$H_{cd}(t)$ drives the system along states $\{\ket{\psi_k(t)}\}$ both in STA applications and in metrology. 
There are, however, some important differences: 

(i) In STA, the states $\ket{\psi_k(t)}$  are eigenstates of a reference Hamiltonian $H_0(t)$ (which plays a similar role than $H_g(t)$ as the Hamiltonian 
whose dynamics we want to transform by adding new terms) while in metrology they are eigenstates of $\partial_gH_g(t)$. 

(ii) In STA, $H_0(t)$ is by construction diagonal in the basis $\{\ket{\psi_k(t)}\}$. In metrology $H_g(t)$ is in general not diagonal in this basis. 

(iii) The functions $f_k(t)$ can be  chosen to simplify the Hamiltonian. They do not produce transitions among the $\{\ket{\psi_k}\}$, they just accumulate a phase factor $e^{-i\theta_k(t)}$ for each $|\psi_k(t)\ra$.
In STA, we may apply this freedom to drive the system along the desired paths with 
$H_{tot}(t)=H_0(t)+H_{cd}(t)$, which is in fact the most common form, instead of $H_{tot}(t)=H_{cd}(t)$. 
By contrast, in metrology we could not in general 
use $H_g(t)+H_{cd}(t)$ because the addition of $H_g(t)$ does more than just changing phases, it produces transitions. 
That is why in metrology $H_{tot}(t)$ is just $H_{cd}(t)$, Equation (\ref{first}), at least as a starting point, {{because}} a reformulation is in fact needed, see  Equation (\ref{htot}) below and the related discussion.  

(iv) In metrology the denomination ``counterdiabatic'' for $H_{cd}$ is, strictly speaking, an abuse of language as, 
in that context, $H_{cd}$ precludes transitions among eigenstates of $\partial_g H_g$, and not transitions among adiabatic states. 
Nevertheless, the formal expressions are identical so that we shall keep the same  terminology and  the 
same notation. 

(v) The emphasis in STA is on fast processes, whereas in the STA-like approach used in metrology speed might be
taken into account but it is not necessarily the main goal. Instead, the emphasis is on a precise parameter estimation.

Let us now come back to metrology. 
To recap, the addition of $H_c$ to $H_g$ would guarantee the state following needed in principle to reach the upper bound, but that is not enough,  there are two very important points to take into account:  

(a) Formally $H_c$ as written above, see Equation (\ref{hc}), depends on $g$, whose exact value is unknown. A way out is to set $H_c$ for an approximate
value $g_c$. 

(b) To get the upper bound of the  Fisher information, in addition to following the {\it state} dynamics, the eigenvalues of $\partial_g H_{tot}$ should be the ``right ones'', i.e., those of $\partial_g H_g$. This point is possibly not fully explicit in \cite{ArticlePang} but it is quite crucial, as the eigenvalues of $\partial_g H_{tot}$
for  $H_{tot}=H_{cd}(g)$ are in general not the right ones. 

The way out to these two points is to reformulate the  control Hamiltonian in (\ref{hc})  as \cite{ArticlePang}
\begin{equation}
\label{generalhc}
H_c(t)|_{g=g_c}=\sum_{k} f_k(t) \ket{\widetilde{\psi}_k(t)} \bra{\widetilde{\psi}_k(t)}-H_{g_c}(t)+i \sum_{k}\ket{\partial_t \widetilde{\psi}_k(t)}\bra{\widetilde{\psi}_k(t)},
\end{equation}
where $\ket{\widetilde{\psi}_k(t)}$ is the $k$th eigenstate of $\partial_g H_g(t)$ with $g=g_c$. In this context, the subscript $g=g_c$ does not mean that the value of $g_c$ is exactly equal to the unknown $g$. Rather, it means that $g_c$ should be written instead of the unknown $g$.

Instead of Equation (\ref{first}),  the total Hamiltonian is thus reformulated as 
\begin{equation}
\label{Htotgeneral}
	H_{tot}(t)=H_g(t)+H_c (t)|_{g=g_c},
\end{equation}
or, taking into account Equation (\ref{HcHcd}),
\begin{equation}
\label{htot}
H_{tot}(t)= H_g(t)-H_g(t)|_{g=g_c}+H_{cd}(t)|_{g=g_c}=H_g(t)-H_{g_c}(t)+H_{cd}(t)|_{g=g_c}.
\end{equation}
This is  finally the structure used in Reference \cite{ArticlePang} to approach the upper bound of the Fisher information. The first term provides the right maximal and minimal eigenvalues of $\partial_g H_g(t)$, 
now $\partial_g H_{tot}(t)=\partial_g H_g(t)$,  whereas the whole sum ($\approx H_{cd}(g)$ but not exactly) 
essentially drives the two corresponding eigenstates as dynamical solutions of the full Hamiltonian.  
This structure implies the need for an ``adaptive scheme'', i.e., a guess value $g_c$ is taken as starting point to produce a better estimate $g_c'$
and so on. Convergence towards $g$ is not guaranteed for arbitrary circumstances, but in specific examples,
the iterations do converge and convergence criteria may be found \cite{ArticlePang}.  
This motivates a further difference between ordinary STA and the STA-like approach: 

(vi) The STA-like approach in metrology is adaptive, it proceeds by iteration to find via measurements, the value $g$.  
In ordinary STA there is no such a scheme, nothing plays the role of successive $g_c, g_c', g_c''...$ values.
There are iterative approaches, 
such as the superadiabatic iterations \cite{ReviewSTA},  but their aim and formal structure do not match closely the described adaptive scheme. Nevertheless, superadiabatic iterations may be the basis for other parameter estimation schemes as sketched in the final discussion.

For the total Hamiltonian (\ref{htot}), Equation (\ref{hgint}) is reformulated  as
\begin{equation}
	h_{g}(T)=\int_{0}^{T}U^{\dagger}(0\rightarrow t)\partial_{g}H_{tot}(t)U(0\rightarrow t)dt,
	\label{hrein}
\end{equation}
where $\partial_{g}H_{tot}=\partial_{g}H_{g}$ by construction;  the unitary evolution operators $U(0\rightarrow t)$ and $U^{\dagger}(0\rightarrow t)$ correspond now to the evolution driven by the total Hamiltonian $H_{tot}$. That is, the reformulated $h_{g}(T)$ depends both on $g$ and $g_c$.

A further remark on notation: We stay essentially faithful to the compact notation in \cite{ArticlePang} 
to facilitate comparison, but compactness comes with a price as we  use some symbols, for example $H_{tot}$ or $h_g$, for different things, contrast in particular 
(\ref{first}) and the reformulation  (\ref{htot}). A more precise  but  heavier notation would likely be cumbersome for the reader. We assume that the context should make clear the right interpretation. Note also that    
in the practical applications of the adaptive method in Reference \cite{ArticlePang} only the reformulated expressions for $H_{tot}$ and $h_g$ are used. In cases where doubts could arise, we will  specify the equation number.

The eigenstates with the maximum and minimum eigenvalues of $\partial_g H_g|_{g=g_c}$  will be denoted as $\ket{ \widetilde{\psi}_{max}(t)}$ and $\ket{ \widetilde{\psi}_{min}(t)}$ respectively.  With the initial state
\begin{equation}
	\ket{\Psi(0)}=\frac{1}{\sqrt{2}}\left(\ket{ \widetilde{\psi}_{max}(0)}+\ket{ \widetilde{\psi}_{min}(0)}\right),
\end{equation}
the maximal Fisher information (\ref{qi}) reaches  in zeroth order in the deviation 
the upper bound (\ref{IQub}).

To attain in practice this upper bound of the quantum Fisher information, the following observable can be measured at time $T$,
\begin{equation}
\label{obs}
	O=\ket{+}\bra{+}-\ket{-}\bra{-},
\end{equation}
where
\begin{equation}
\ket{\pm}=\frac{1}{\sqrt{2}}\left(e^{-i\theta_{max}}\ket{ \widetilde{\psi}_{max}(T)} \pm e^{-i\theta_{min}}\ket{ \widetilde{\psi}_{min}(T)} \right).
\end{equation}
Keeping dominant orders in $\delta g=g-g_c$,   
\begin{eqnarray}
	\langle O \rangle 
	&=&\cos(\delta g \int_0^T(\mu_{max}(t)-\mu_{min}(t))dt),
	\end{eqnarray}
\begin{equation}
\label{O2}
	\langle \Delta O^2 \rangle= \sin[2](\delta g \int_0^T(\mu_{max}(t)-\mu_{min}(t))dt),
\end{equation}
so $g$ can be found from the estimator $\langle O \rangle$. The variance of the estimate is the inverse of the upper bound of the Fisher information,  
\begin{equation}
\label{deltag2}
\delta g^2=\frac{\langle \Delta O^2 \rangle}{|\partial_{\delta g}\langle O \rangle|^2}=\frac{1}{\left[\int_0^T(\mu_{max}(t)-\mu_{min}(t))dt\right]^2}.
\end{equation}
\section{Estimation of Field Amplitude and Rotation Frequency}
\label{chap:estimation}
Pang and Jordan \cite{ArticlePang} apply the above methodology  to  a qubit in a uniformly rotating magnetic field $\mathbf{B}(t)=B[\cos(\omega t)\mathbf{e}_x+\sin(\omega t)\mathbf{e}_z]$, where $\mathbf{e}_x$ and $\mathbf{e}_z$ are unit vectors with directions  $\hat{x}$ and $\hat{z}$, respectively, to estimate  the amplitude $B$ and the rotation frequency $\omega$. 
Here, we shall focus on $\omega$ as it leads to the most interesting results. 

The Hamiltonian that represents the interaction between the qubit and the field is
\begin{equation}
\label{sysham}
H_\omega(t)=-B[\cos({\omega t})\, \sigma_x+\sin({\omega t})\, \sigma_z],  
\end{equation}
in terms of Pauli matrices.
{We may as well consider  a reinterpretation of this 
Hamiltonian as the semiclassical interaction for a two level system in a properly set laser or microwave field, but let us keep formally the notation for a magnetic field.}

An interesting exercise is to compute the Fisher information for  $\omega$ (now the $g$ parameter),  with and without Hamiltonian control.  
The derivative of $H_\omega$ is  
\begin{equation}
\label{partialHw}
	\partial_{\omega}H_{\omega}(t)=t B[\sin(\omega t)\sigma_x-\cos(\omega t)\sigma_z],
\end{equation}
with time-dependent eigenvalues  $\mu_{max,min}=\pm tB$. Using this result in Equation (\ref{IQub}), the upper bound of the Fisher information is
\begin{equation}
\label{Iwmax}
I_{\omega~ub}^{(Q)}=B^2 T^4.
\end{equation}
This result is nontrivial  because  the gap between Hamiltonian eigenvalues is not increased. Otherwise, if the Hamiltonian is set to increase rapidly with time, arbitrarily high powers of $T$ or  exponential growth may be found \cite{ArticlePang}. Note also that the maximum power of $T$ that can be achieved with a time independent Hamiltonian is $T^2$.

Without Hamiltonian control the optimal quantum Fisher information {{(\ref{IQtau})}} is instead 
\begin{equation}
\label{IQ0}
	I_{\omega,0}^{(Q)}\sim \frac{4B^2T^2}{4B^2+\omega^2}.
\end{equation}

\subsection{Estimation of the Rotation Frequency with Hamiltonian Control}

If we assume $f_k(t)=0$, (\ref{generalHcd}) 
becomes
\begin{equation}
H_{cd}=-\frac{\omega}{2}\sigma_y.
\end{equation}
Note that $H_{cd}$  is here a time-independent Hamiltonian with an upper bound $\sim T^2$ for the Fisher information. This 
illustrates the general statement made before that $H_{cd}$ drives along the ``right'' eigenvectors of $\partial_g H_g$ but does not 
necessarily provide the right eigenvalues as $\partial_g H_{cd}\ne \partial_g H_{g}$.      
 
As explained in Section \ref{sec:pang}, the way out is to reformulate $H_c(t)$ at the estimated value $\omega_c$
and set 
%
\begin{eqnarray}
\label{Htotwc}
H_{tot}&=& H_{\omega}(t)+H_c(t)|_{\omega=\omega_c}
\nonumber\\
\label{ht}
&=&-B[\cos({\omega t})\sigma_x+\sin({\omega t})\sigma_z]+B[\cos({\omega_c t})\sigma_x+\sin({\omega_c t})\sigma_z]-\frac{\omega_c}{2}\sigma_y.
\end{eqnarray}
To compute the optimal Fisher information,  the  corresponding $h_{\omega}(T)$ in Equation (\ref{hrein}) is diagonalized  to find 
its eigenvalues. 
 Since  $\omega_c$ is assumed close to $\omega$, $h_{\omega}(T)$ is expanded around $\omega_c=\omega$ as
\begin{equation}
\label{expansion}
h_{\omega}(T) = h_{\omega}(T)|_{\omega_c=\omega}+\partial_{\omega_c}h_{\omega}(T)|_{\omega_c=\omega}\delta \omega+\frac{1}{2!}\partial_{\omega_c}^2h_{\omega}(T)|_{\omega_c=\omega}\delta \omega^2+\dots,
\end{equation}
where $\delta \omega=\omega_c-\omega$, (Notice that this notation in \cite{ArticlePang} is not consistent with $\delta g =g-g_c$.) 
\begin{equation}
\label{hwfinal}
	h_{\omega}(T)=-\frac{BT^2}{2}\sigma_z-\frac{BT^3}{3}\sigma_x \delta \omega+\left(\frac{4B^2T^5}{15}+\frac{BT^4}{4}\sigma_z\right)\frac{\delta \omega^2}{2}+\mathcal{O}(\delta \omega^3),
\end{equation}
with eigenvalues 
\begin{equation}
\label{taumaxmin}
	\tau_{max,min}=\pm\frac{BT^2}{2}\mp \frac{BT^4}{72}\delta \omega^2 +\mathcal{O}(\delta \omega^4).
\end{equation}
Therefore, substituting the results given by Equation (\ref{taumaxmin}) into Equation (\ref{IQtau}), the optimal Fisher information 
for (\ref{Htotwc}) becomes
\begin{eqnarray}
 	I_{\omega}^{(Q)}
 	&=&
B^2T^4 \left(1-\frac{1}{18}T^2\delta\omega^2\right),
\label{IQw}
 \end{eqnarray}
where higher order terms of $\delta \omega$ have been neglected.
Conditions for convergence are analyzed in 
\cite{ArticlePang}.

\section{Alternative Driving via Physical Unitary Transformations}
\label{Chapters:Unitary}

In ordinary applications of STA based on the counterdiabatic approach,  $H_{cd}$ often implies different operators from those in the reference Hamiltonian $H_0$. These extra operators  may be hard or even impossible to generate in the laboratory. In the applications to metrology of $STA$-like methods the same 
difficulties may arise with the control Hamiltonian $H_c$. 
Specifically, for the system and Hamiltonian studied in Section \ref{chap:estimation}, the control Hamiltonian includes a $\sigma_y$ term whose implementation could be quite challenging in some systems \cite{Bason2011}, this really depends on the particular realization
of the two-level system, but here we shall assume, as a basic exercise, that $\sigma_y$ is a term that we want to avoid. 
In STA applications, it is sometimes possible to change the structure of the total Hamiltonian avoiding undesired terms by means of 
``physical'' unitary transformations \cite{Ibanez2012_100403,Martinez-Garaot2014_053408,ReviewSTA}.  We shall explore this approach in the context of  parameter estimation. 
Specifically our generic goal is  to modify the total Hamiltonian $H_{tot}$ (see Equation (\ref{htot})) so that we get rid of the problematic 
terms. In the example of the previous section we will modify Equation (\ref{ht}), to get rid of the $\sigma_y$ term without losing the $T^4$ dependence in the Fisher information. 

Given a Hamiltonian $H(t)$ that drives the general state $\ket{\psi(t)}$, the unitarily transformed state $\ket{\psi'(t)}=G^{\dagger}(t)\ket{\psi(t)}$ obeys
\begin{equation}
i\partial_t \ket{\psi'(t)}=H'(t)\ket{\psi'(t)},
\end{equation}
where
\begin{eqnarray}
\label{HI}
H'(t)&=&G^{\dagger}(t)[H(t)-K(t)]G(t),\\
\label{K}
K(t)&=&i \dot{G}(t) G^{\dagger}(t),
\end{eqnarray}
and the dot stands for time derivative. $H'(t)$ is in general not just the unitary transform of $H(t)$ when $G(t)$ depends on time. Notice also that although these expressions are formally the same as those that define an interaction picture, here the alternative Hamiltonian $H'(t)$ and $H(t)$ represent different physical drivings just like $\ket{\psi(t)}$ and $\ket{\psi'(t)}$ represent different dynamic states. In the context of STA methods, the transformation provides indeed an alternative shortcut to the one represented by $H$ if we set
\begin{eqnarray} 
\label{cond1}
G(0)&=&G(T)=1, \\
\label{cond2}
\dot{G}(0)&=&\dot{G}(T)=0,
\end{eqnarray}
in order to guarantee
\begin{eqnarray} 
\ket{\psi'(0)}=\ket{\psi(0)},\;\;\; 
\ket{\psi'(T)}=\ket{\psi(T)},
\end{eqnarray}
and
\begin{eqnarray} 
H'(T)=H(T),\;\;\;
H'(0)=H(0). 
\end{eqnarray}
That is, with these boundary conditions the wavefunctions and the Hamiltonians coincide at the boundary times.
In ordinary STA, these boundary conditions may be relaxed in some cases
\cite{Martinez-Garaot2014_053408}. Moreover, in metrology as we shall see.  

Let us now examine the operator $h'_g=i  U'^\dagger_g \partial_g U'_g$ corresponding to $H'$, 
where  
\beqa
\label{UG}
U'_g&=& G^\dagger U_g. 
\eeqa
The parameter $g$ is unknown so we assume that the unitary transformation $G$ does not depend on it.  Then
\begin{equation}
h'_g(T)=i  U'^\dagger_g \partial_g U'_g
= i  U^\dagger_g \partial_g U_g =h_g(T).
\end{equation}
Similarly   $h'^2_g(T)=h_g^2(T)$, 
so $H$ and $H'$ will have the same maximal Fisher information (four times the variance of $h_g$, see Equation (\ref{FI})) for the same 
initial state. The optimal and upper bound Fisher information depend only on $h'_g(T)=h_g(T)$ so they are also {{unaffected}}.  
In this context, 
there is no need in principle for the transformation operator $G$ to satisfy the boundary conditions in Equations (\ref{cond1}) and (\ref{cond2}). 
However, it may be convenient to satisfy Equation (\ref{cond1}) so that the wavefunctions $|\Psi(t)\ra$ and $|\Psi'(t)\ra$ coincide at both initial and final times. 
In particular this would allow us to use the same observable $O$ in Equation (\ref{obs}) as an estimator for $g$.

In the example of Section \ref{chap:estimation}, we  want to transform the Hamiltonian
(\ref{ht}) 
to get rid of $\sigma_y$. 
We really need  the full Hamiltonian (\ref{ht}) as starting point. If we instead used a pure 
$H_{cd}=-\frac{\omega}{2}\sigma_y$, as in Equation (\ref{first}), only a 
$T^2$ dependence for the Fisher information could be reached since this Hamiltonian is time-independent.   

When the Hamiltonian is a linear combination of generators of some Lie algebra, the unitary transformation $G$ may be constructed by exponentiating elements of the algebra and imposing the vanishing of the unwanted terms \cite{Martinez-Garaot2014_053408}. In 
our example, the generators of the algebra are the Pauli matrices so, we will choose unitary transformations of the form 
\begin{equation}
\label{unitrans}
G(t)=e^{-i\alpha(t)\sigma_i},
\end{equation}
where $\alpha(t)$ is a given time-dependent real function and $\sigma_i$ can be any of the Pauli matrices $\{\sigma_x$, $\sigma_y$, $\sigma_z\}$. 
Taking into account that
\begin{eqnarray}
\label{properties}
e^{i\alpha(t)\sigma_i}\sigma_ie^{-i\alpha(t)\sigma_i}&=&\sigma_i, \nonumber \\
e^{i\alpha(t)\sigma_i}\sigma_je^{-i\alpha(t)\sigma_i}&=&\cos[2\alpha(t)]\sigma_j+\frac{i}{2}\sin[2\alpha(t)][\sigma_i,\sigma_j],~~~~~~~i\neq j,
\end{eqnarray}
we choose $\sigma_i=\sigma_y$ in Equation (\ref{unitrans}).  
The alternative Hamiltonian becomes
\beqa
\label{HpG}
H'(t)=&-&B\left[\cos{(\omega t+2\alpha)}-\cos{(\omega_c t+2\alpha)} \right] \sigma_x \nonumber \\
&-&\left[ \frac{\omega_c}{2}+\dot\alpha \right] \sigma_y \nonumber \\
&-&B\left[\sin{(\omega t+2\alpha)}-\sin{(\omega_c t+2\alpha)} \right] \sigma_z. 
\eeqa
To cancel the $\sigma_y$ term, we choose
\begin{equation}
\label{alphagen}
\alpha(t)=-\frac{\omega_ct}{2}.
\end{equation}
{ (A similar unitary transformation 
is also used in \cite{ArticlePang} with different aim and results, see the Appendix \ref{appa1}
for further details.)}
Substituting Equation (\ref{alphagen}) into Equation (\ref{HpG}) he have finally
\begin{equation}
\label{HpGa}
H'(t)= B\left \{ 1-\cos{[(\omega-\omega_c)t]}\right \}\sigma_x
-B \sin{[(\omega-\omega_c)t]}\sigma_z,
\end{equation}
which has the same structure (generators) as the reference Hamiltonian (\ref{sysham}) but different time-dependent coefficients.
Rewriting (\ref{HpGa}) as
\beqa
H'(t)&=&B \left [ 1- \cos{(\omega t)}\cos{(\omega_c t)}-\sin{(\omega t)}\sin{(\omega_c t)}\right ] \sigma_x \nonumber \\
&-&B \left [ \cos{(\omega_c t)} \sin{(\omega t)} -\cos{(\omega t)} \sin{(\omega_c t)}\right ] \sigma_z,
\eeqa
it can be seen that the realization of $H'(t)$ is possible assuming that fields  oscillating  with $\omega$ (a ``carrier'' signal with precise frequency to be determined) and $\omega_c$ (a test signal with accurately known frequency)
can be implemented and combined.  Alternatively, we may think of a setting where the difference between two frequencies $\omega-\omega_c$ can  be controlled accurately even if the carrier frequency is unknown. 
Therefore, the alternative feasible Hamiltonian will keep the $T^4$ scaling of Fisher information for a given evolution time $T$ and, consequently, the estimation of the $\omega$ will be the same than the one achieved using the untransformed Hamiltonian. 
Specifically an explicit perturbative calculation in orders of $\delta \omega$ reproduces the result in Equation (\ref{IQw}) in agreement with the general proof   
given above that the unitary transformation does not change the Fisher information. 

As for the observable $O$ in Equation (\ref{obs}), it may be used as an estimator provided $G(0)=G(T)=1$. Then the final states 
for drivings by $H$ and $H'$ are identical if the initial states are the same. Equation (\ref{alphagen}) implies that $G(0)=1$.  
Noting that, generically,  
\begin{equation}
\label{exponential}
\exp[\pm i \eta(t) \sigma_i]=\cos[\eta(t)] I\pm i \sin[\eta(t)]\sigma_i,\\ \\
\end{equation}
at periodic times 
\beq
T=2n 2\pi/\omega_c,
\eeq
$G(T)=1$ as desired.

\section{Discussion}
The seminal work of Pang and Jordan in \cite{ArticlePang} demonstrates that 
time-dependent Hamiltonians allow for better parameter estimations than time-independent ones. Specifically, the time dependence of the Fisher information can be given by higher powers of time without increasing Hamiltonian intensity. In the example of a qubit  in a rotating magnetic field, the optimal Fisher information for the rotating frequency of the field can reach a $T^4$ dependence surpassing the limit $T^2$ of time-independent Hamiltonians. 
In practice, it is 
necessary in general to add a control Hamiltonian to reach the upper bound of the Fisher information.  
Pang and Jordan propose a control Hamiltonian 
to reach the upper bound using an  STA-like adaptive approach. 

We have discussed  similarities and differences between actual STA methods and the  STA-like method in Reference \cite{ArticlePang}.  This analysis sets the ground  to apply other STA techniques  in metrology. We have explored here one of them: physical 
(rather than formal) unitary transformations.  
We have first proven that for these transformations  
the Fisher information does not change.  Then, a ``proof of principle'' application is worked out for the frequency measurement in the single qubit model:
assuming that a $\sigma_y$ type of term is not easy to implement, as it happens, e.g., in  the experimental setting in 
\cite{Bason2011},  we  find, by a unitary transformation,  alternative Hamiltonians leading to the upper bound of the  Fisher information without a $\sigma_y$ term.

As for further possibilities, we sketch here some ideas to be developed in more 
detail elsewhere: 
The counterdiabatic approach may be regarded  as the zeroth iteration of an STA-generating scheme based on  superadiabatic 
iterations \cite{Berry1987,Demirplak2008,Ibanez2013,ReviewSTA}. 
In zeroth order a given (Schr\"odinger picture) Hamiltonian $H_0(t)$ is diagonilized with a basis $\{|n_0(t)\ra\}$ to set an interaction picture (IP)
based on the unitary transformation $A_0=\sum_n|n_0(t)\ra\la n_0(0)|$ with IP Hamiltonian  $H_1=A_0^\dagger (H_0-K_0) A_0$, where 
$K_0=i\dot A_0 A_0^\dagger$. If, in the IP, the coupling term is cancelled by adding its negative, $A_0^\dagger K_0 A_0$, the dynamics unfolds without transitions. Back in the Schr\"odinger picture (SP) this amounts to driving the system aided by a counterdiabatic term,  with the modified Hamiltonian $H_0+K_0$, where $K_0=H_{cd}=H_{cd}^{(0)}$. 
The added superindex $(0)$ denotes that higher order iterations may be worked out by repeating the same process, starting, in the first iteration, with 
$H_1$ instead of $H_0$. This first ``superadiabatic'' iteration generates a different coupling term that may be canceled with its 
negative as in the  CD method. Of course further iterations could be implemented. The different iteration-dependent uncoupling terms $H_{cd}^{(j)}$ added to $H_0$ in the SP may or may not be useful depending on three main points \cite{Ibanez2013}: (a) their ``intensity'' does not necessarily decrease with the iteration, typically it decreases first until it begins to grow \cite{Berry1987}; (b) the operators involved in $H_{cd}^{(j)}$---in some operator basis---change with the iteration. Thus, some iterations may lead to undesired terms, difficult to implement, or, instead, to a convenient operator structure; (c) the higher order iterations, $j\ge 1$, beyond the CD method do not necessarily drive the system from $|n_0(0)\ra$ to a final $|n_0(T)\ra$, the eigenvectors of the original Hamiltonian, up to phase factors. For that end some  boundary conditions should be satisfied, namely,    
that the successive $K_j$ vanish at the time boundaries \cite{Ibanez2013}. In STA these conditions are satisfied by Hamiltonians $H_0$ with vanishing successive time derivatives \cite{Garrido1964,Ibanez2013}.  

To apply the above scheme to parameter estimation several changes are needed. 
First of all, the states $|n_0(t)\ra$ should represent eigenstates of $\partial_g H_g(t)$ instead of eigenstates of the reference Hamiltonian $H_g$. Then, the SP for  a given iteration 
cannot be $H_g+H_{cd}^{(j)}$ because $H_g$ is now not  diagonal in  $\{|n_0(t)\ra\}$ in general. In addition, the fact that $g$ is not accurately known should be taken into account.   
An adaptive reformulation is therefore needed similar to the one in \cite{ArticlePang}. Thus for an iteration $j$, instead of $H_g+H_{cd}^{(j)}$,  the reformulated 
SP Hamiltonian  must be of the form 
\beq
H_g-H|_{g=g_c}+H_{cd}^{(j)}|_{g=g_c},
\eeq
and, moreover, the boundary conditions for the $K_j$ should be satisfied.     

One more point to consider is that in STA,  the spectral information of $H_0$  (eigenstates and eigenvectors) may not 
be easily available if at all. This is often the case in many-body systems. 
Therefore, building $H_{cd}$ with the usual recipe is not possible so 
several approximate techniques have been worked out that do not use spectral information
\cite{Sels2017,Claeys2019,Petiziol2018,Petiziol2019}.
These techniques might be applicable in parameter estimation protocols 
when the spectral information of 
$\partial_g H_g$ is not accessible.    
{ Other interesting open questions are the analysis of mixed, rather than pure, states \cite{Fiderer2019},  
and looking for possible connections between the bound in Quantum Fisher information 
and quantum speed limits \cite{Campbell2018}.}  
\vspace*{.2cm}\\   

\section*{Acknowledgement}
This work was supported by the Basque Country Government (Grant No. IT986-16), by   PGC2018-101355- B-I00 (MCIU/AEI/FEDER,UE).

\begin{appendix}
\section{Physical vs. Formal Transformations} 
\label{appa1}
In this Appendix, we compare the ``interaction picture'' transformation 
in the Supplementary Note 3 of \cite{ArticlePang} and the transformation in Section \ref{Chapters:Unitary}.  While in both cases  similar unitary operators,  $\exp(-i \sigma_y \omega t/2)$ and $\exp(i \sigma_y \omega_c t/2)$, are  applied, the aim, results,  and physical content of the transformations are not the same. In 
\cite{ArticlePang}, the two Hamiltonians involved are  
(\ref{sysham})
and 
\beq
-B \sigma_x+\omega \sigma_y/2.
\label{tindep}
\eeq
Formally, (\ref{sysham}) may be regarded as  an interaction picture Hamiltonian of  the ``Schr\"odinger'' Hamiltonian  (\ref{tindep}) 
if the interaction picture wavefunction is defined in terms of the Schr\"odinger picture one as 
$\psi_{IP}=\exp(i\omega t\sigma_y/2)\psi_S$. As in ordinary applications of interaction pictures,     
the physics is the same in both pictures, they are just different representations of the same thing, and the aim of the transformation 
is to get a simple expression of the evolution operator driven by (\ref{sysham}) making use of the  time independent structure of (\ref{tindep}). In these equations only the exact value $\omega$ appears, and no control Hamiltonian has been added. 

In Section \ref{Chapters:Unitary}, the starting Hamiltonian is instead (\ref{ht}), which is transformed via 
(\ref{HI}) using $G=\exp(i\omega_ct\sigma_y/2)$ into (\ref{HpGa}). The two Hamiltonians involved, (\ref{ht}) 
and (\ref{HpGa}),  are different now from those in Reference \cite{ArticlePang}, (\ref{sysham}) and (\ref{tindep}).         
 Moreover, the distinction between $\omega$ and $\omega_c$ plays a fundamental role, and the control Hamiltonian 
is added in (\ref{ht}).  In Section \ref{Chapters:Unitary}, the two related Hamiltonians  represent different physical settings and drive different dynamics. The transformation is now made to change the physics, it is not just a convenient, 
formal change of representation. 
\end{appendix} 

\label{Bibliography}
\bibliography{MyCollection}
\end{document}